\documentclass[12pt]{article}
\usepackage{cjp}

\begin{document}
\begin{flushright}
{UT-Komaba 99-18}
\end{flushright}
\vskip 0.5 truecm

\title{Gauged Gross--Neveu model with overlap
 fermions\footnote{Talk by K.Nagao at Chiral'99, Taipei, Taiwan,
 Sep.13-18, 1999.}}
\author{Ikuo Ichinose and  Keiichi Nagao}  
\address{Institute of Physics, University of Tokyo, Komaba,
  Tokyo, 153-8902 Japan } 
\date{\today} 
\maketitle
\begin{abstract}
We investigate chiral properties of the overlap lattice
fermion by using solvable model in two dimensions,
the gauged Gross-Neveu model.
In this model, the chiral symmetry is spontaneously broken
in the presence of small but finite fermion mass.
We calculate the quasi-Nambu-Goldstone(NG) boson mass 
as a function of the bare fermion mass
and two parameters in the overlap formula.
We find that the quasi-NG boson
mass has desired properties as a result of 
the extended chiral symmetry found by L\"uscher.
We also show the PCAC relation is satisfied
in desired form.
Comparison between the overlap and Wilson lattice fermions  
is also made.
\end{abstract}

\section{Introduction}
Species doubling is a long standing problem in the lattice fermion
formulation.
Wilson fermion is the most suitable formulation\cite{Wilsonterm} and it is used
in most of the numerical studies of lattice gauge theory.
However in order to reach the desired continuum limit, fine tuning
must be done with respect to the ``bare fermion mass" and the 
Wilson parameter.
Recently a very promising formulation of lattice fermion
named overlap fermion was proposed by Narayanan and Neuberger\cite{overlap}.
In that formula the Ginsparg and Wilson(GW)
 relation\cite{GinspargWilson} plays a very important
role, and because of that there exists an ``extended" (infinitesimal)
chiral symmetry.
In this paper we shall study or test the overlap fermion by
using the gauged Gross-Neveu model in two dimensions.
This is a solvable model which has similar
chiral properties with QCD$_4$, i.e., chiral symmetry is spontaneously
broken with a small but finite bare fermion mass and pion appears
as quasi-Nambu-Goldstone boson.
Actually a closely related model was studied on a lattice in order to test
properties of the Wilson fermion in the continuum limit\cite{Ichinose}.
Therefore advantage of the overlap fermion becomes clear
by the investigation in this paper.

In order to examine chiral properties of the overlap fermion, we define
lattice Gross-Neveu model in two ways.
\begin{itemize}
\item 
In the first one, the interaction terms respect the ordinary chiral
symmetry and break the extended chiral symmetry. 
Then it is not obvious if quasi-massless pions appear (in the continuum 
limit ) with or without any finite tuning of the parameters in the 
overlap fermion.
\item
On the other hand, the second one respect the extended chiral symmetry
and as a result the interaction terms become nonlocal.
Then it is not clear if the Goldstone theorem is applicable in this system. 
\end{itemize}
All these questions are answered in this paper.
The first question is important for a formulation of lattice theory
with a symmetry which cannot be realized exactly on a lattice 
like supersymmetry.
The second one is important for study of QCD with exact extended 
chiral symmerty.
Especially study of the second lattice model shows what is an order
parameter for the extended chiral symmetry and which field behaves
as Nambu-Goldstone boson.

\section{The first model}
The first model is defined by the following action on 
lattice with the lattice spacing $a$,
\begin{eqnarray}
S&=& {N \over 2} \sum_{pl} \prod U_\mu(n) 
+ a^2\sum_{n,m}\bar{\psi}(m)D(m,n)\psi(n)
+a^2M_B\sum_n\bar{\psi}\psi(n)   \nonumber \\
&& 
-{a^2 \over \sqrt{N}}\sum_n\Big[\phi^i(n)(\bar{\psi}\tau^i\psi)(n)
+\phi^i_5(n)(\bar{\psi}\tau^i\gamma_5\psi)(n)\Big] \nonumber  \\
&& +{a^2 \over 2g_v}\sum_n\Big[\phi^i(n)\phi^i(n)+
\phi^i_5(n)\phi^i_5(n)\Big],
\label{action1}
\end{eqnarray}
where $U_\mu(n)$ is U(1) gauge field defined on links,
$\psi^l_{\alpha}\; (\alpha=1,...,N,
l=1,...,L)$ are fermion fields with flavour index $l$,
and the matrix $\tau^i\; (i=0,...,L^2-1)$ acting on the flavour
index is normalized as
\begin{equation}
{\rm{Tr}}(\tau^i\tau^k)=\delta_{ik}
\end{equation}
and 
\begin{equation}
\tau^0={1 \over \sqrt{L}}, \; \; \{\tau^i,\tau^j\}=d^{ijk}\tau^k,
\end{equation}
where $d^{ijk}$'s are the structure constants of $SU(L)$.
Fields $\phi^i$ and $\phi^i_5$ are scalar and pseudo-scalar
bosons, respectively.
The covariant derivative in Eq.(\ref{action1}) is defined by the
overlap formula
\begin{eqnarray}
D&=&{1\over a}\Big(1+X{1 \over \sqrt{X^{\dagger}X}}\Big),  \nonumber  \\
X_{nm}&=&\gamma_{\mu}C_{\mu}(n,m)+B(n,m),  \nonumber   \\
C_{\mu}&=&{1 \over 2a}\Big[\delta_{m+\mu,n}U_{\mu}(m)-
\delta_{m,n+\mu}U^{\dagger}_{\mu}(n)\Big],  \nonumber  \\
B(n,m)&=&-{M_0\over a}+{r\over 2a}\sum_{\mu}\Big[2\delta_{n,m}
-\delta_{m+\mu,n}U_{\mu}(m)-\delta_{m,n+\mu}U^{\dagger}_{\mu}(n)\Big],
\label{covD}
\end{eqnarray} 
where $r$ and $M_0$ are 
dimensionless nonvanishing free parameters of the overlap lattice fermion 
formalism\cite{overlap,Neuberger}.
The overlap Dirac operator $D$ does not have the ordinary chiral 
invariance but satisfies the GW relation instead,
\begin{equation}
D\gamma_5+\gamma_5D=aD\gamma_5D.
\label{GWr}
\end{equation}
From (\ref{action1}) it is obvious that the systematic $1/N$ expansion
is possible and we shall employ it.

The action (\ref{action1}) contains the bare fermion mass $M_B$
which explicitly breaks the chiral symmetry.
This bare mass also breaks the following infinitesimal 
transformation, which was discovered by L\"uscher\cite{Luscher} and we call
``extended chiral symmetry",
\begin{eqnarray}
&& \psi(n) \rightarrow \psi(n)+\tau^k\theta^k\gamma_5\Big\{
\delta_{nm}-aD(n,m)\Big\}\psi(m),  \nonumber  \\
&& \bar{\psi}(n) \rightarrow \bar{\psi}(n)+\bar{\psi}(n)
\tau^k\theta^k\gamma_5
  \nonumber  \\
&&\phi^i(n) \rightarrow \phi^i(n)+d^{ikj}\theta^k\phi^j_5(n),  \nonumber  \\
&&\phi^i_5(n) \rightarrow \phi^i_5(n)-d^{ikj}\theta^k\phi^j(n),
\label{extended}
\end{eqnarray}
where $\theta^i$ is an infinitesimal transformation parameter.
It is also verifed that the interaction terms in (\ref{action1})
also explicitly break the extended chiral symmetry.

From the action (\ref{action1}), it is obvious that 
$\phi^i$ and $\phi^i_5$ are composite fields
of the fermions,
\begin{equation}
\phi^i={g_v \over \sqrt{N}}\bar{\psi}\tau^i\psi, \;\;
\phi^i_5={g_v \over \sqrt{N}}\bar{\psi}\gamma_5\tau^i\psi.
\label{phi}
\end{equation}
As in the continuum model, we expect that the field
$\phi^0$ acquires a nonvanishing vacuum expectation value(VEV),
\begin{equation}
\langle\phi^0\rangle=\sqrt{NL}M_s,
\label{VEV}
\end{equation}
and we define subtracted fields,
\begin{eqnarray}
&&\varphi^0=\phi^0-\sqrt{NL}M_s,  \nonumber  \\
&&\varphi^i=\phi^i\; \; (i\neq 0),  \;\; \varphi^i_5=\phi^i_5.
\end{eqnarray}

From the chiral symmetry and (\ref{VEV}), one may expect that 
quasi-Nambu-Goldstone(NG) bosons appear as a result of the spontaneous
breaking of the chiral symmetry.
They are nothing but $\varphi^i_5$.
However as we explained above, this expectation can {\em not} be
accepted straightforwardly because of explicit breaking of 
both the ordinary and extended chiral symmetries.
Careful studies are then required.

The VEV $M_s$ is  determined by the tadpole cancellation
condition of $\varphi^0$.
In order to perform an explicit calculation of the $1/N$-expansion,
we introduce the gauge potential $\lambda_{\mu}(n)$ in the usual way,
i.e., $U(n,\mu)=\exp ({ia\over \sqrt{N}}\lambda_{\mu}(n))$
and employ the weak-coupling expansion by Kikukawa 
and Yamada\cite{weakexpansion}.

The effective action of the pion and the gauge boson is obtained by 
integrating over the quark fields.
For details, see Ref.\cite{IN}.
The pion part is given as follows in the leading-order of $1/N$,
\begin{equation}
S^{(2)}_{eff}[\varphi_5]=\int_k{1 \over 2}
\varphi^i_5(-k)\Gamma^5_{ij}(k^2)\varphi^j_5(k)
\label{Svphi}
\end{equation}
where 
\begin{eqnarray}
\Gamma^5_{ij}(k^2)&=&\delta_{ij}\Big[{1 \over g_v}+
\int_k {\rm{Tr}}[\gamma_5\langle \psi(k-p)\bar{\psi}(k-p)\rangle
\gamma_5\langle \psi(k)\bar{\psi}(k)\rangle]\Big]  \nonumber  \\
&=&\delta_{ij}\Big[\epsilon+2k^2M_0^2 A(k^2;M)\Big],\label{Gamma5} \\
M&=&M_B+M_s.
\end{eqnarray}
Parameter $\epsilon$ in $\Gamma^5_{ij}$ is proportional
to the pion mass and measures the derivation from the limit of the exact
chiral symmetry.
Practical calculation gives 
\begin{equation}
\epsilon ={M_BM_0^2  \over M_s}\Big[ -\ln (M_0M^2a)+\mbox{const.}\Big]
+O(a).
\label{epsilon}
\end{equation}
Therefore $\epsilon \propto M_B+O(a)$ and the limit $M_B \rightarrow 0$ 
is considered as the chiral limit.
This result is in sharp contrast with the Wilson fermion.
There fine tunning of the the ``bare mass" $M_{W,B}$
and the Wilson parameter $r_W$ is required in order to reach
the chiral limit\cite{Ichinose}.

It is also straightforward to calculate $A(k^2;M)$ in (\ref{Gamma5}).
In the continuum limit,
\begin{equation}
A(k^2;M)\rightarrow {1 \over 4\pi\mu^2}+O(k^2).
\label{Ak}
\end{equation}
where $\mu=M_0M$ 
and therefore the pion mass is given as $m^2_\pi=2\pi M^2\epsilon$.

There exists a mixing term of the gauge boson $\lambda_\mu$ and 
the pion $\varphi^0_5$,
\begin{equation}
S^{(2)}_{eff}[\lambda_\mu,\varphi^0_5] =-2\sqrt{L}
M^2_0M\int_k\sum \lambda_\mu(-k)
\epsilon_{\mu\nu}k_\nu A(k^2;M)\varphi^0_5(k),
\label{mixing}
\end{equation}
which is identical with the continuum calculation.
This mixing term is related to the discussion of the  U(1) problem 
in QCD$_4$ and the above result suggests that the correct anomaly 
appears in the Ward-Takahashi identity of the axial-vector current.

We shall examine the PCAC relation.
By changing variables
as follows in the path-integral representation of 
the partition function,
\begin{eqnarray}
&& \psi(n) \rightarrow \psi(n)+\tau^k\theta^k(n)\gamma_5\Big\{
\delta_{nm}-aD(n,m)\Big\}\psi(m),  \nonumber  \\
&& \bar{\psi}(n) \rightarrow \bar{\psi}(n)\left\{1+\tau^k\theta^k(n)\gamma_5
\right\},  \nonumber  \\
&&\phi^i(n) \rightarrow \phi^i(n)+d^{ikj}\theta^k\phi^j_5(n),  \nonumber  \\
&&\phi^i_5(n) \rightarrow \phi^i_5(n)-d^{ikj}\theta^k\phi^j(n),
\label{extended2}
\end{eqnarray}
we obtain the Ward-Takahashi(WT) identity,
\begin{eqnarray}
&&\langle \partial_\mu j^k_{5,\mu}(n)-2M(\bar{\psi}\tau^k\gamma_5\psi)(n)
+{2\sqrt{N}\over g_v}M_s\varphi^k_5(n)  \nonumber  \\
&& \; \; \; \;  +D^k_A(n)-\delta^{k0}N\sqrt{L}a
{\rm{Tr}}[\gamma_5D(n,n)]\rangle=0,
\label{WT}
\end{eqnarray}
where the last term comes from the measure of the path integral,
and the explicit form of the current operator $j^k_{5,\mu}$ is 
obtained by Kikukawa and Yamada\cite{axialvector}.
The above WT identity is expressed in terms of the pions and gauge
boson by integrating over the quarks.
We obtain the final form of the WT identity {\em in the continuum
limit},
\begin{eqnarray}
\partial_\mu j^k_{5,\mu}&=&i\delta^{k0}{\sqrt{NL} \over \pi}
\sum_{\mu\nu}\epsilon_{\mu\nu}\partial_\nu\lambda_\mu+2M\epsilon\sqrt{N}
\varphi^k_5  \nonumber   \\
&=&i\delta^{k0}{\sqrt{NL} \over \pi}
\sum_{\mu\nu}\epsilon_{\mu\nu}\partial_\nu\lambda_\mu+\sqrt{{2N\over \pi}}
m^2_\pi\times {\varphi^k_5\over \sqrt{2\pi M^2}}.
\label{WT2}
\end{eqnarray}
Then it is obvious that the PCAC relation is satisfied in the
overlap fermion formalism.

\section{The second model}
We shall turn to the second lattice Gross-Neveu model.
We modify the interaction terms as follows,
\begin{eqnarray}
S_2&=& -{N \over 2} \sum_{pl} \prod U_\mu(n) 
+ a^2\sum_{n,m}\bar{\psi}(m)D(m,n)\psi(n)
-a^2M_B\sum_n\bar{\psi}\psi(n)   \nonumber \\
&& 
-{a^2 \over \sqrt{N}}\sum_n\Big[\phi^i(n)\bar{\psi}(n)\tau^i
\Big(\delta_{nm}-{a\over 2}D(n,m)\Big)\psi(m)  \nonumber  \\
&&  +\phi^i_5(n)\bar{\psi}(n)\tau^i\gamma_5
\Big(\delta_{nm}-{a\over 2}D(n,m)\Big)\psi(m)\Big] \nonumber  \\
&& +{a^2 \over 2g_v}\sum_n\Big[\phi^i(n)\phi^i(n)+
\phi^i_5(n)\phi^i_5(n)\Big].
\label{actionII}
\end{eqnarray}
It is straightforward to verify the invariance of the action $S_2$
under the transformation (\ref{extended}).
From the action (\ref{actionII}), $\phi^i$ and $\phi^i_5$
are {\em nonlocal} composite fields of the fermions in the present case,
\begin{equation}
\phi^i={g_v\over \sqrt{N}}\bar{\psi}\tau^i\Big(1-{a \over 2}D\Big)
\psi, \;\; 
\phi^i_5={g_v\over \sqrt{N}}\bar{\psi}\tau^i\gamma_5\Big(1-{a \over 2}D\Big)
\psi.
\label{eq.motion}
\end{equation}
As in the previous case we expect that the field
$\phi^0$ acquires a nonvanishing vacuum expectation value (VEV)
\begin{equation}
\langle\phi^0\rangle=\sqrt{NL}M_s,
\label{VEV2}
\end{equation}
and we define subtracted fields.
In Ref.\cite{Chandra}, it is argued that the nonlocal
composite field $\phi^0$ works as an order parameter
for the extended chiral symmetry.

The effective action of the pions and gauge boson is obtained
by a similar method as before.
For detailes, see Ref.\cite{IN2}.
Especially the pion mass is obtained at {\em finite lattice
spacing} as follows,
\begin{equation}
\epsilon \propto {M_Ba^2 \over M_s}.
\label{epsilon2}
\end{equation}
This expression (\ref{epsilon2}) should be compared with 
(\ref{epsilon}).
In the second lattice model with the exact extended chiral symmetry,
the Goldstone theorem holds and the (pion mass)$^2 \propto M_B$
at finite lattice spacing.
This result is welcome but it should be remarked that this result
is obtained in the weak-coupling (or large $N$) region.
We also verified that the PCAC relation appears in the
desired form at finite lattice spacing.

\bigskip
{\bf Acknowledgments}  \\
KN would like to thank the organizers of Chiral'99, especially,
Prof.T.W.Chiu for his warm hospitality.

\newpage

 %
 %
 %

\end{document}